\documentclass{article}

\usepackage[english]{babel}

\usepackage[letterpaper,top=2cm,bottom=2cm,left=3cm,right=3cm,marginparwidth=1.75cm]{geometry}

\usepackage{amsmath}
\usepackage{amssymb}
\usepackage{graphicx}
\usepackage{indentfirst}
\usepackage{tikz}
\usepackage{tikz-3dplot}
\usepackage{pgfplots}
 \pgfplotsset{compat=1.18}
\usepackage{url}
\usepackage{subcaption}
\usepackage[numbers]{natbib}
\usepackage{comment}
\usepgfplotslibrary{groupplots}
\usetikzlibrary{positioning}

\title{DeFi Liquidation Risk Modeling Using Geometric Brownian Motion}
\author{Timofei Belenko, Georgii Vosorov}
\date{}

\begin{document}
\maketitle

\begin{abstract}
In this paper, we propose an analytical method to compute the collateral liquidation probability in decentralized finance (DeFi) stablecoin single-collateral lending. Our approach models the collateral exchange rate as a zero-drift geometric Brownian motion, and derives the probability of it crossing the liquidation threshold. Unlike most existing methods that rely on computationally intensive simulations such as Monte Carlo, our formula provides a lightweight, exact solution. This advancement offers a more efficient alternative for risk assessment in DeFi platforms.
\end{abstract}

\section{Introduction}

In recent years, decentralized finance (DeFi) lending platforms have gained significant popularity as innovative alternatives to traditional financial services. These platforms enable users to borrow and lend digital assets in a transparent manner. One of the key concerns in DeFi lending is the risk of collateral liquidation, which can cause substantial financial losses for lenders and borrowers alike. Collateral liquidation occurs when the collateral value falls below the liquidation threshold. Therefore, developing robust methods to assess and quantify the likelihood of liquidation events is essential to ensure the stability and security of DeFi ecosystems.

Predicting liquidation events on DeFi lending platforms is inherently challenging due to the high volatility of cryptocurrency prices.

Furthermore, existing methods for estimating liquidation probabilities often rely on computationally intensive techniques such as Monte Carlo simulations. Although these approaches can provide valuable information, they tend to be resource intensive and time consuming, limiting their practicality for real-time risk assessment and decision making. Developing analytical solutions that are accurate and computationally efficient remains an important goal in the field.

Modeling collateral exchange rates as zero-drift Wiener processes offers a significant advantage in terms of computational efficiency and accuracy. This approach allows for the derivation of explicit analytical formulas, eliminating the need for resource-intensive simulations, such as Monte Carlo methods. As a result, the solution becomes not only faster to compute, but also more precise, providing reliable estimates of liquidation probabilities without the errors typically associated with numerical simulations. This makes the method particularly suitable for real-time risk assessment and decision making in the volatile environment of cryptocurrency markets.

\section{Exploratory Data Analysis}

After collecting and organizing information on the daily changes in the opening and closing exchange rates of a selected cryptocurrency, we conducted an initial exploratory analysis to better understand its behavior over time. One of the first visualizations we produced was a histogram representing the distribution of daily rate changes (Figure \ref{histogram}).

From the shape of the histogram, it is evident that the data approximately follows a bell-shaped curve, resembling the normal distribution. This suggests that most of the daily changes cluster around the mean, with fewer extreme values occurring at the tails.

\begin{figure}
    \centering
\begin{tikzpicture}
    \begin{axis}[
        title = Exchange Rate Change Close $-$ Open Distribution for Period: 29.03.2024 - 29.03.2025,
        width=\textwidth,
        height=23cm,
        xmin=-400, xmax=600,
        ymin=0, ymax=40,
        xtick distance=200,
        ytick distance=1,
        minor y tick num = 0,
        area style,
    ]
    \addplot+[ybar interval] plot coordinates { (-380, 1) (-360, 0) (-320, 2) (-300, 1) (-280, 3) (-260, 1) (-240, 4) (-220, 4) (-200, 2) (-180, 6) (-160, 6) (-140, 10) (-120, 6) (-100, 16) (-80, 25) (-60, 28) (-40, 23) (-20, 33) (0, 31) (20, 39) (40, 33) (60, 31) (80, 16) (100, 9) (120, 6) (140, 6) (160, 6) (180, 4) (200, 1) (220, 2) (240, 3) (260, 0) (300, 2) (320, 1) (340, 1) (360, 1) (380, 0) (580, 1) (600, 0)};
\end{axis}
\end{tikzpicture}
\caption{Histogram of daily change in Ethereum exchange rate}
\label{histogram}
\end{figure}

\section{Deriving Analytical Solution}

\subsection{Liquidation Criteria}

Liquidation occurs when the value of the collateral drops below the liquidation threshold. Collateral value changes due to exchange rates, interest rates and the loan value, although exchange rate volatility diminishes the relevance of interest rates, allowing them to be disregarded in the context of mathematical modeling. We will model loans of stablecoins; therefore, we can assume that the loan value is constant and only model the collateral exchange rates. In addition, we will model single-collateral loans.

\subsection{Modeling Collateral Exchange Rates as Zero-Drift Geometric Brownian Motion}

A classic way to quantitatively model financial assets is Geometric Brownian Motion (GBM) \cite{Topcu_Guloksuz_2021} \cite{Stojkoski_2020}. It also fits the collected data. We will use it to model our exchange rates. It has the following stochastic differential equation (SDE):
$$dS_t = \sigma S_t dW_t$$
where $\sigma$ is the standard deviation or volatility, $S_t$ is the exchange rate, and $W_t$ is a Wiener process.
We can rewrite it as:
$$dS_t = x\left(t, S_t\right)dt + y\left(t, S_t \right)dW_t$$
where $x\left(t, S_t\right) = 0$ and $y\left(t, S_t \right) = \sigma S_t$. We can define $f\left(Y_t\right) = ln\left(Y_t\right)$ and now we can apply It\^{o} lemma to get:
$$df = \left( \frac{\partial f}{\partial t} + x \frac{\partial f}{\partial S_t} + \frac{y^2}{2}\frac{\partial^2 f}{\partial S_t^2} \right) dt + y \frac{\partial f}{\partial S_t} dW_t$$
Since $f\left(Y_t\right) = ln\left(Y_t\right)$, $\frac{\partial f}{\partial t} = 0$, $\frac{\partial f}{\partial S_t} = \frac{1}{S_t}$ and $\frac{\partial^2 f}{\partial S_t^2} = - \frac{1}{S_t^2}$
After simplifying we get:
$$d\left(ln\left(S_t\right)\right) = df = \sigma dW_t - \frac{\sigma^2}{2}dt$$
We can integrate from $0$ to $t$ over $dt$ and get:
$$ln\left(Y_t\right) - ln\left(Y_0\right) = \sigma W_t - \frac{\sigma^2}{2}t$$
Define $X_t = ln\left(Y_t\right)$:
$$X_t = X_0 + \sigma W_t - \frac{\sigma^2}{2}t$$
which is a drifting Brownian motion.
For small periods of time or small volatility, drift can be assumed to be zero, and approximation can be achieved via the reflection principle. Otherwise, we will have to use the inverse Gaussian distribution.

\subsection{Applying the Reflection Principle}

Since we disregard interest rates, we can define the liquidation threshold and the exchange rate at which the collateral value reaches the liquidation threshold to be constant. Therefore, we can use the reflection principle.

Currently, we will disregard the drift $-\frac{\sigma^2}{2}t$ in the GBM equation to approximate a solution as we said earlier. Later, we will see for what time periods this approximation works.

The reflection principle tells us about the probability that a Wiener process will cross some value in some time \cite{lalley_brownian}.
Here it is:
$$\mathbb{P} \left(  \underset{0 \leq s \leq t}{inf} W_s \leq a \right) = 2 \mathbb{P} \left( W_t \leq a \right) $$

To ensure applicability to $X$, we can modify it. The starting exchange rate is $S_0$, so $X_0 = ln(S_0)$. Similarly, $X_{liquidation} = ln(S_{liquidation})$. We can now define $a$:
$$a = X_{\text{liquidation}}-X_0=ln \left(S_{\text{liquidation}} \right) - ln \left( S_0 \right) = ln \left( \frac{S_{\text{liquidation}}}{S_0} \right)$$
After introducing standard deviation $\sigma$ we get:
$$\mathbb{P}\left(\text{liquidation}\right) = \mathbb{P}\left(\underset{0 \leq s \leq t}{inf} X_s \leq a\right) = 2 \mathbb{P}\left(X_t \leq a\right) = 2 \Phi\left( \frac{a}{\sigma \sqrt{t}} \right) = 2 \Phi \left( \frac{ln \left( \frac{S_{\text{liquidation}}}{S_0} \right)}{\sigma \sqrt{t}} \right)$$
Where $\Phi$ is the cumulative distribution function (CDF) of standard normal distribution. Now we have a complete analytical solution to our problem.
We have derived the closed-form solution for our approximation.

\subsection{Applying the Inverse Gaussian Distribution}

The inverse Gaussian distribution $IG\left(\mu_0, \lambda\right)$ can be thought as the distribution of the first hitting time of a threshold $a>0$ for Brownian motion\cite{Folks_Inverse_Gaussian}:
$$
\begin{array}{l}
    X_0=0 \\
    X_t=\sigma W_t + \mu t \; , \; \mu>0
\end{array}
$$
We can ensure $X_0 = 0$ by shifting the coordinates to fit $X_0 = 0$. We get $a_1 = X_\text{liquidation} - X_0 = ln\left(S_\text{liquidation}\right) - ln\left(S_0\right) = ln\left(\frac{S_\text{Liquidation}}{S_0}\right) < 0$. We know $\mu_1 = -\frac{\sigma^2}{2} < 0$. These are negative but have the same sign. Now, flip the coordinate system, and we get $a_2 = -a_1 = ln\left(\frac{S_0}{S_\text{Liquidation}}\right) > 0$ and $\mu_2 = -\mu_1 = \frac{\sigma^2}{2}$. The first passage time probability density function (PDF) of such process has the following formula:
$$f\left(t_1\right) = \frac{a_2}{\sigma\sqrt{2\pi t^3}} exp\left(-\frac{\left(a_2 - \mu_2 t\right)^2}{2 \sigma^2 t}\right)$$
for
$$t_1=inf\left\{ \tau>0 \; | \; X_\tau = a_2\right\} \sim IG\left(\frac{a_2}{\mu_2}, \; \left(\frac{a_2}{\sigma}\right)^2\right)$$
We want to predict the probability of it crossing the threshold in time $t$ from the start.
Since the PDF of the inverse Gaussian distribution is $0$ for negative inputs, we only need to compute the CDF for $t$. CDF for $IG\left(\mu_0, \lambda\right)$ at the value $x$ is:
$$F\left(x\right) = \Phi\left(\sqrt{\frac{\lambda}{x}}\left(\frac{x}{\mu_0} - 1\right)\right) + exp\left(\frac{2\lambda}{\mu_0}\right)\Phi\left(-\sqrt{\frac{\lambda}{x}}\left(\frac{x}{\mu_0} + 1\right)\right)$$
Replacing $x$ with $t$, $\mu_0$ with $\frac{a_2}{\mu_2}$ and $\lambda$ with $\left(\frac{a_2}{\sigma}\right)^2$ results in the probability of liquidation.

\section{Simulating Collateral Exchange Rates}

Monte Carlo simulation \cite{glasserman_mc} \cite{higham2001} \cite{kloeden_platen} can be used to validate the theoretical formula derived earlier. To simulate collateral exchange rate paths, we model the evolution of the price as a stochastic process using a small time step $dt$. There are two common ways to discretize the GBM for simulation:

\begin{itemize}
    \item \textbf{Euler–Maruyama approximation:} This linear method updates the exchange rate using the first-order approximation of the SDE, assuming
    $$dS \sim \sigma S_t \sqrt{dt} Z,\quad Z \sim \mathcal{N}\left(0,1\right)$$
    which leads to:
    $$S_{t+dt} = S_t + \sigma S_t \sqrt{dt} Z$$
    
    \item \textbf{Exact GBM solution:} Alternatively, we can use the closed-form solution to the GBM SDE, which yields multiplicative updates:
    $$S_{t+dt} = S_t \exp\left( \sigma \sqrt{dt} Z - \frac{\sigma^2}{2}dt\right)$$

    \item \textbf{Zero-drift GBM solution:} Since we have neglected drift for the reflection principle approximation. We will also simulate it:
    $$S_{t+dt} = S_t \exp\left( \sigma \sqrt{dt} Z\right)$$
    
\end{itemize}

For each simulated price path, we check whether the price crosses the liquidation threshold at any time. The simulated liquidation probability is then calculated as:
$$\mathbb{P}(\text{liquidation})_{\text{Monte Carlo}} = \frac{N_{\text{liquidated}}}{N}$$
where $N_{\text{liquidated}}$ is the number of paths that were liquidated and $N$ is the total number of simulated paths.

\section{Simulation Results and Validation}

\begin{table}[]
\begin{tabular}{|c|c|c|c|c|}
\hline
Row & Annualized Volatility & Starting Exchange Rate & Liquidation Exchange Rate & $\frac{S_0}{S_\text{liquidation}}$ \\
\hline
1   & 0.8                   & 1500                   & 1200                      & 1.25                               \\
2   & 0.8                   & 2500                   & 1200                      & 2.083333                           \\
3   & 1.8                   & 1500                   & 1200                      & 1.25                               \\
4   & 1.8                   & 2500                   & 1200                      & 2.083333                           \\
5   & 2.8                   & 1500                   & 1200                      & 1.25                               \\
6   & 2.8                   & 2500                   & 1200                      & 2.083333                           \\
\hline
\end{tabular}
\caption{Simulation values}
\label{tab:values}
\end{table}

\begin{figure}
    \centering
\begin{tikzpicture}

    \begin{groupplot}[
        group style={
            group size=3 by 6, 
            horizontal sep=0.75cm, 
            vertical sep=1cm
        },
        height=3.5cm, 
        width=5.5cm   
    ]

    \nextgroupplot[legend to name=beta]
    \addplot[red] coordinates {(3.0, 0.002093)
 (7.0, 0.043994)
 (14.0, 0.154383)
 (30.0, 0.33059)
 (60.0, 0.491476)
 (90.0, 0.574307)
 (120.0, 0.626639)
 (150.0, 0.663485)
 (180.0, 0.691223)};\addlegendentry{Reflection Principle};
    \addplot[blue] coordinates {(3.0, 0.002339)
 (7.0, 0.04913)
 (14.0, 0.172266)
 (30.0, 0.368359)
 (60.0, 0.546548)
 (90.0, 0.637673)
 (120.0, 0.694862)
 (150.0, 0.734864)
 (180.0, 0.764778)};\addlegendentry{Inverse Gaussian};
    \addplot[green] coordinates {(3.0, 0.0009)
 (7.0, 0.0304)
 (14.0, 0.1364)
 (30.0, 0.3177)
 (60.0, 0.5)
 (90.0, 0.5973)
 (120.0, 0.6622)
 (150.0, 0.6962)
 (180.0, 0.7359)};\addlegendentry{Monte Carlo - GBM};
    \addplot[yellow] coordinates {(3.0, 0.0012)
 (7.0, 0.0263)
 (14.0, 0.1129)
 (30.0, 0.2806)
 (60.0, 0.4501)
 (90.0, 0.52)
 (120.0, 0.5985)
 (150.0, 0.6369)
 (180.0, 0.6703)};\addlegendentry{Monte Carlo - No Drift GBM};
    \addplot[black] coordinates {(3.0, 0.002)
 (7.0, 0.0342)
 (14.0, 0.1341)
 (30.0, 0.3167)
 (60.0, 0.5022)
 (90.0, 0.5964)
 (120.0, 0.6614)
 (150.0, 0.7054)
 (180.0, 0.7405)};\addlegendentry{Monte Carlo - Euler-Maruyama};
 \coordinate (top) at (rel axis cs:0,1);
    \nextgroupplot[legend to name=alpha, scaled y ticks=base 10:2]
    \addplot[color = lime] coordinates {(3.0, 0.001193)
 (7.0, 0.013594)
 (14.0, 0.017983)
 (30.0, 0.01289)
 (60.0, -0.008524)
 (90.0, -0.022993)
 (120.0, -0.035561)
 (150.0, -0.032715)
 (180.0, -0.044677)};\addlegendentry{Reflection Principle $-$ GBM};
    \addplot[color = cyan] coordinates {(3.0, 0.000893)
 (7.0, 0.017694)
 (14.0, 0.041483)
 (30.0, 0.04999)
 (60.0, 0.041376)
 (90.0, 0.054307)
 (120.0, 0.028139)
 (150.0, 0.026585)
 (180.0, 0.020923)};\addlegendentry{Reflection Principle $-$ No Drift GBM};
    \addplot[color = orange] coordinates {(3.0, 9.3e-05)
 (7.0, 0.009794)
 (14.0, 0.020283)
 (30.0, 0.01389)
 (60.0, -0.010724)
 (90.0, -0.022093)
 (120.0, -0.034761)
 (150.0, -0.041915)
 (180.0, -0.049277)};\addlegendentry{Reflection Principle $-$ Euler-Maruyama};
    \addplot[color=pink] coordinates {(0,0)};\addlegendentry{Inverse Gaussian $-$ GBM};
    \addplot[color=teal] coordinates {(0,0)};\addlegendentry{Inverse Gaussian $-$ Euler-Maruyama};
    \nextgroupplot[scaled y ticks=base 10:2]
    \addplot[color = pink] coordinates {(3.0, 0.001439)
 (7.0, 0.01873)
 (14.0, 0.035866)
 (30.0, 0.050659)
 (60.0, 0.046548)
 (90.0, 0.040373)
 (120.0, 0.032662)
 (150.0, 0.038664)
 (180.0, 0.028878)};
    \addplot[color = teal] coordinates {(3.0, 0.000339)
 (7.0, 0.01493)
 (14.0, 0.038166)
 (30.0, 0.051659)
 (60.0, 0.044348)
 (90.0, 0.041273)
 (120.0, 0.033462)
 (150.0, 0.029464)
 (180.0, 0.024278)};

    \nextgroupplot
    \addplot[red] coordinates {(3.0, 0.0)
 (7.0, 0.0)
 (14.0, 3e-06)
 (30.0, 0.001373)
 (60.0, 0.023644)
 (90.0, 0.064657)
 (120.0, 0.109579)
 (150.0, 0.152384)
 (180.0, 0.191394)};
    \addplot[blue] coordinates {(3.0, 0.0)
 (7.0, 0.0)
 (14.0, 4e-06)
 (30.0, 0.001971)
 (60.0, 0.033779)
 (90.0, 0.092003)
 (120.0, 0.155368)
 (150.0, 0.215353)
 (180.0, 0.269663)};
    \addplot[green] coordinates {(3.0, 0.0)
 (7.0, 0.0)
 (14.0, 0.0)
 (30.0, 0.0016)
 (60.0, 0.0271)
 (90.0, 0.0834)
 (120.0, 0.1365)
 (150.0, 0.2046)
 (180.0, 0.2471)};
    \addplot[yellow] coordinates {(3.0, 0.0)
 (7.0, 0.0)
 (14.0, 0.0)
 (30.0, 0.0007)
 (60.0, 0.0181)
 (90.0, 0.0578)
 (120.0, 0.0978)
 (150.0, 0.1397)
 (180.0, 0.1747)};
    \addplot[black] coordinates {(3.0, 0.0)
 (7.0, 0.0)
 (14.0, 0.0)
 (30.0, 0.0017)
 (60.0, 0.0299)
 (90.0, 0.0836)
 (120.0, 0.1436)
 (150.0, 0.1988)
 (180.0, 0.2553)};
    \nextgroupplot[scaled y ticks=base 10:2]
    \addplot[lime] coordinates {(3.0, 0.0)
 (7.0, 0.0)
 (14.0, 3e-06)
 (30.0, -0.000227)
 (60.0, -0.003456)
 (90.0, -0.018743)
 (120.0, -0.026921)
 (150.0, -0.052216)
 (180.0, -0.055706)};
    \addplot[cyan] coordinates {(3.0, 0.0)
 (7.0, 0.0)
 (14.0, 3e-06)
 (30.0, 0.000673)
 (60.0, 0.005544)
 (90.0, 0.006857)
 (120.0, 0.011779)
 (150.0, 0.012684)
 (180.0, 0.016694)};
    \addplot[orange] coordinates {(3.0, 0.0)
 (7.0, 0.0)
 (14.0, 3e-06)
 (30.0, -0.000327)
 (60.0, -0.006256)
 (90.0, -0.018943)
 (120.0, -0.034021)
 (150.0, -0.046416)
 (180.0, -0.063906)};
    \nextgroupplot[scaled y ticks=base 10:2]
    \addplot[pink] coordinates {(3.0, 0.0)
 (7.0, 0.0)
 (14.0, 4e-06)
 (30.0, 0.000371)
 (60.0, 0.006679)
 (90.0, 0.008603)
 (120.0, 0.018868)
 (150.0, 0.010753)
 (180.0, 0.022563)};
    \addplot[teal] coordinates {(3.0, 0.0)
 (7.0, 0.0)
 (14.0, 4e-06)
 (30.0, 0.000271)
 (60.0, 0.003879)
 (90.0, 0.008403)
 (120.0, 0.011768)
 (150.0, 0.016553)
 (180.0, 0.014363)};


    \nextgroupplot
    \addplot[red] coordinates {(3.0, 0.171498)
 (7.0, 0.370692)
 (14.0, 0.526743)
 (30.0, 0.665442)
 (60.0, 0.759786)
 (90.0, 0.802856)
 (120.0, 0.828827)
 (150.0, 0.846661)
 (180.0, 0.859876)};
    \addplot[blue] coordinates {(3.0, 0.19134)
 (7.0, 0.412875)
 (14.0, 0.585417)
 (30.0, 0.73698)
 (60.0, 0.837602)
 (90.0, 0.881984)
 (120.0, 0.90786)
 (150.0, 0.925041)
 (180.0, 0.937354)};
    \addplot[green] coordinates {(3.0, 0.1211)
 (7.0, 0.3147)
 (14.0, 0.4968)
 (30.0, 0.6674)
 (60.0, 0.7978)
 (90.0, 0.8443)
 (120.0, 0.8843)
 (150.0, 0.9038)
 (180.0, 0.9224)};
    \addplot[yellow] coordinates {(3.0, 0.1026)
 (7.0, 0.2831)
 (14.0, 0.4431)
 (30.0, 0.5898)
 (60.0, 0.7084)
 (90.0, 0.7643)
 (120.0, 0.7896)
 (150.0, 0.8085)
 (180.0, 0.828)};
    \addplot[black] coordinates {(3.0, 0.1295)
 (7.0, 0.3177)
 (14.0, 0.4993)
 (30.0, 0.666)
 (60.0, 0.7909)
 (90.0, 0.8501)
 (120.0, 0.8792)
 (150.0, 0.908)
 (180.0, 0.9232)};
    \nextgroupplot[scaled y ticks=base 10:2]
    \addplot[lime] coordinates {(3.0, 0.050398)
 (7.0, 0.055992)
 (14.0, 0.029943)
 (30.0, -0.001958)
 (60.0, -0.038014)
 (90.0, -0.041444)
 (120.0, -0.055473)
 (150.0, -0.057139)
 (180.0, -0.062524)};
    \addplot[cyan] coordinates {(3.0, 0.068898)
 (7.0, 0.087592)
 (14.0, 0.083643)
 (30.0, 0.075642)
 (60.0, 0.051386)
 (90.0, 0.038556)
 (120.0, 0.039227)
 (150.0, 0.038161)
 (180.0, 0.031876)};
    \addplot[orange] coordinates {(3.0, 0.041998)
 (7.0, 0.052992)
 (14.0, 0.027443)
 (30.0, -0.000558)
 (60.0, -0.031114)
 (90.0, -0.047244)
 (120.0, -0.050373)
 (150.0, -0.061339)
 (180.0, -0.063324)};
    \nextgroupplot[scaled y ticks=base 10:2]
    \addplot[pink] coordinates {(3.0, 0.07024)
 (7.0, 0.098175)
 (14.0, 0.088617)
 (30.0, 0.06958)
 (60.0, 0.039802)
 (90.0, 0.037684)
 (120.0, 0.02356)
 (150.0, 0.021241)
 (180.0, 0.014954)};
    \addplot[teal] coordinates {(3.0, 0.06184)
 (7.0, 0.095175)
 (14.0, 0.086117)
 (30.0, 0.07098)
 (60.0, 0.046702)
 (90.0, 0.031884)
 (120.0, 0.02866)
 (150.0, 0.017041)
 (180.0, 0.014154)};


    \nextgroupplot
    \addplot[red] coordinates {(3.0, 7e-06)
 (7.0, 0.003235)
 (14.0, 0.037339)
 (30.0, 0.154939)
 (60.0, 0.314551)
 (90.0, 0.411552)
 (120.0, 0.476991)
 (150.0, 0.52473)
 (180.0, 0.561475)};
    \addplot[blue] coordinates {(3.0, 1e-05)
 (7.0, 0.004639)
 (14.0, 0.053264)
 (30.0, 0.218921)
 (60.0, 0.438419)
 (90.0, 0.567411)
 (120.0, 0.651561)
 (150.0, 0.710929)
 (180.0, 0.755125)};
    \addplot[green] coordinates {(3.0, 0.0)
 (7.0, 0.0029)
 (14.0, 0.0352)
 (30.0, 0.1904)
 (60.0, 0.4043)
 (90.0, 0.5338)
 (120.0, 0.6344)
 (150.0, 0.688)
 (180.0, 0.725)};
    \addplot[yellow] coordinates {(3.0, 0.0)
 (7.0, 0.0017)
 (14.0, 0.0281)
 (30.0, 0.1314)
 (60.0, 0.2816)
 (90.0, 0.375)
 (120.0, 0.4511)
 (150.0, 0.4902)
 (180.0, 0.5345)};
    \addplot[black] coordinates {(3.0, 0.0)
 (7.0, 0.0051)
 (14.0, 0.04)
 (30.0, 0.1909)
 (60.0, 0.4044)
 (90.0, 0.5386)
 (120.0, 0.6275)
 (150.0, 0.6905)
 (180.0, 0.737)};
    \nextgroupplot[scaled y ticks=base 10:2]
    \addplot[lime] coordinates {(3.0, 7e-06)
 (7.0, 0.000335)
 (14.0, 0.002139)
 (30.0, -0.035461)
 (60.0, -0.089749)
 (90.0, -0.122248)
 (120.0, -0.157409)
 (150.0, -0.16327)
 (180.0, -0.163525)};
    \addplot[cyan] coordinates {(3.0, 7e-06)
 (7.0, 0.001535)
 (14.0, 0.009239)
 (30.0, 0.023539)
 (60.0, 0.032951)
 (90.0, 0.036552)
 (120.0, 0.025891)
 (150.0, 0.03453)
 (180.0, 0.026975)};
    \addplot[orange] coordinates {(3.0, 7e-06)
 (7.0, -0.001865)
 (14.0, -0.002661)
 (30.0, -0.035961)
 (60.0, -0.089849)
 (90.0, -0.127048)
 (120.0, -0.150509)
 (150.0, -0.16577)
 (180.0, -0.175525)};
    \nextgroupplot[scaled y ticks=base 10:2]
    \addplot[pink] coordinates {(3.0, 1e-05)
 (7.0, 0.001739)
 (14.0, 0.018064)
 (30.0, 0.028521)
 (60.0, 0.034119)
 (90.0, 0.033611)
 (120.0, 0.017161)
 (150.0, 0.022929)
 (180.0, 0.030125)};
    \addplot[teal] coordinates {(3.0, 1e-05)
 (7.0, -0.000461)
 (14.0, 0.013264)
 (30.0, 0.028021)
 (60.0, 0.034019)
 (90.0, 0.028811)
 (120.0, 0.024061)
 (150.0, 0.020429)
 (180.0, 0.018125)};


    \nextgroupplot
    \addplot[red] coordinates {(3.0, 0.379376)
 (7.0, 0.564973)
 (14.0, 0.684067)
 (30.0, 0.781028)
 (60.0, 0.844171)
 (90.0, 0.872494)
 (120.0, 0.889458)
 (150.0, 0.901065)
 (180.0, 0.909646)};
    \addplot[blue] coordinates {(3.0, 0.422507)
 (7.0, 0.627435)
 (14.0, 0.75708)
 (30.0, 0.859674)
 (60.0, 0.922678)
 (90.0, 0.94868)
 (120.0, 0.963032)
 (150.0, 0.972086)
 (180.0, 0.978259)};
    \addplot[green] coordinates {(3.0, 0.2848)
 (7.0, 0.503)
 (14.0, 0.6657)
 (30.0, 0.8102)
 (60.0, 0.8845)
 (90.0, 0.9267)
 (120.0, 0.9474)
 (150.0, 0.9629)
 (180.0, 0.9691)};
    \addplot[yellow] coordinates {(3.0, 0.24)
 (7.0, 0.4406)
 (14.0, 0.582)
 (30.0, 0.7007)
 (60.0, 0.7876)
 (90.0, 0.829)
 (120.0, 0.8493)
 (150.0, 0.8652)
 (180.0, 0.8705)};
    \addplot[black] coordinates {(3.0, 0.2897)
 (7.0, 0.4833)
 (14.0, 0.6619)
 (30.0, 0.7946)
 (60.0, 0.8871)
 (90.0, 0.9236)
 (120.0, 0.9456)
 (150.0, 0.9566)
 (180.0, 0.9682)};
    \nextgroupplot[scaled y ticks=base 10:2]
    \addplot[lime] coordinates {(3.0, 0.094576)
 (7.0, 0.061973)
 (14.0, 0.018367)
 (30.0, -0.029172)
 (60.0, -0.040329)
 (90.0, -0.054206)
 (120.0, -0.057942)
 (150.0, -0.061835)
 (180.0, -0.059454)};
    \addplot[cyan] coordinates {(3.0, 0.139376)
 (7.0, 0.124373)
 (14.0, 0.102067)
 (30.0, 0.080328)
 (60.0, 0.056571)
 (90.0, 0.043494)
 (120.0, 0.040158)
 (150.0, 0.035865)
 (180.0, 0.039146)};
    \addplot[orange] coordinates {(3.0, 0.089676)
 (7.0, 0.081673)
 (14.0, 0.022167)
 (30.0, -0.013572)
 (60.0, -0.042929)
 (90.0, -0.051106)
 (120.0, -0.056142)
 (150.0, -0.055535)
 (180.0, -0.058554)};
    \nextgroupplot[scaled y ticks=base 10:2]
    \addplot[pink] coordinates {(3.0, 0.137707)
 (7.0, 0.124435)
 (14.0, 0.09138)
 (30.0, 0.049474)
 (60.0, 0.038178)
 (90.0, 0.02198)
 (120.0, 0.015632)
 (150.0, 0.009186)
 (180.0, 0.009159)};
    \addplot[teal] coordinates {(3.0, 0.132807)
 (7.0, 0.144135)
 (14.0, 0.09518)
 (30.0, 0.065074)
 (60.0, 0.035578)
 (90.0, 0.02508)
 (120.0, 0.017432)
 (150.0, 0.015486)
 (180.0, 0.010059)};


    \nextgroupplot
    \addplot[red] coordinates {(3.0, 0.003836)
 (7.0, 0.058377)
 (14.0, 0.18075)
 (30.0, 0.360541)
 (60.0, 0.517934)
 (90.0, 0.597574)
 (120.0, 0.64755)
 (150.0, 0.682611)
 (180.0, 0.708943)};
    \addplot[blue] coordinates {(3.0, 0.005499)
 (7.0, 0.083112)
 (14.0, 0.254879)
 (30.0, 0.500105)
 (60.0, 0.702602)
 (90.0, 0.796952)
 (120.0, 0.851525)
 (150.0, 0.886827)
 (180.0, 0.911281)};
    \addplot[green] coordinates {(3.0, 0.0035)
 (7.0, 0.0546)
 (14.0, 0.206)
 (30.0, 0.4455)
 (60.0, 0.6692)
 (90.0, 0.7706)
 (120.0, 0.8372)
 (150.0, 0.8666)
 (180.0, 0.8949)};
    \addplot[yellow] coordinates {(3.0, 0.0018)
 (7.0, 0.0378)
 (14.0, 0.142)
 (30.0, 0.307)
 (60.0, 0.4701)
 (90.0, 0.5545)
 (120.0, 0.606)
 (150.0, 0.6518)
 (180.0, 0.6751)};
    \addplot[black] coordinates {(3.0, 0.0059)
 (7.0, 0.0689)
 (14.0, 0.2125)
 (30.0, 0.4473)
 (60.0, 0.6603)
 (90.0, 0.7691)
 (120.0, 0.8301)
 (150.0, 0.871)
 (180.0, 0.8993)};
    \coordinate (bot) at (rel axis cs:0,-0.9);
    \nextgroupplot[scaled y ticks=base 10:2]
    \addplot[lime] coordinates {(3.0, 0.000336)
 (7.0, 0.003777)
 (14.0, -0.02525)
 (30.0, -0.084959)
 (60.0, -0.151266)
 (90.0, -0.173026)
 (120.0, -0.18965)
 (150.0, -0.183989)
 (180.0, -0.185957)};
    \addplot[cyan] coordinates {(3.0, 0.002036)
 (7.0, 0.020577)
 (14.0, 0.03875)
 (30.0, 0.053541)
 (60.0, 0.047834)
 (90.0, 0.043074)
 (120.0, 0.04155)
 (150.0, 0.030811)
 (180.0, 0.033843)};
    \addplot[orange] coordinates {(3.0, -0.002064)
 (7.0, -0.010523)
 (14.0, -0.03175)
 (30.0, -0.086759)
 (60.0, -0.142366)
 (90.0, -0.171526)
 (120.0, -0.18255)
 (150.0, -0.188389)
 (180.0, -0.190357)};
    \nextgroupplot[scaled y ticks=base 10:2]
    \addplot[pink] coordinates {(3.0, 0.001999)
 (7.0, 0.028512)
 (14.0, 0.048879)
 (30.0, 0.054605)
 (60.0, 0.033402)
 (90.0, 0.026352)
 (120.0, 0.014325)
 (150.0, 0.020227)
 (180.0, 0.016381)};
    \addplot[teal] coordinates {(3.0, -0.000401)
 (7.0, 0.014212)
 (14.0, 0.042379)
 (30.0, 0.052805)
 (60.0, 0.042302)
 (90.0, 0.027852)
 (120.0, 0.021425)
 (150.0, 0.015827)
 (180.0, 0.011981)};

    \end{groupplot}
    \path (top)--(bot) coordinate[midway] (group center);
    \node[above,rotate=90] at (group center -| current bounding box.west) {Results};
    \node[right=-17.25em,inner sep=0pt] at(bot -| current bounding box.south) {\pgfplotslegendfromname{beta}};
    \node[right=-0.61em,inner sep=0pt] at(bot -| current bounding box.south) {\pgfplotslegendfromname{alpha}};
\end{tikzpicture}
\caption{Simulation results and comparison of methods. Probability of liquidation over the time frame in days. Simulation values can be found in table \ref{tab:values}. 10000 paths were simulated for each method with a timestep of 1 day.}
\label{fig:simulation}
\end{figure}

Figure \ref{fig:simulation} shows the simulation results. As expected, the formula derived from the reflection principle works well for low volatility and smaller time frames, but performs poorly under high volatility. The difference between the GBM and Euler-Maruyama simulations and the formula also grows with time. However, the difference between the formula and the no-drift GBM simulation stays consistent. We also see that the formula derived from the inverse Gaussian distribution generally shows more accurate results.

\section{Applications in DeFi Protocol Design}

One of the most direct and impactful applications of our analytical liquidation model lies in its integration with protocols like Aave, one of the leading DeFi lending platforms. In Aave \cite{aave-liquidations} \cite{aave-risks}, each borrowing position is associated with a health factor, which determines how close it is to liquidation. This health factor is based on the current loan-to-value (LTV) ratio, asset prices, and collateral thresholds. However, it does not explicitly represent the probability that a position will be liquidated within a given time frame. By incorporating our closed-form liquidation probability, Aave could offer an additional dimension to risk assessment: a forward-looking probability metric that captures expected liquidation risk over the next N days, based on historical volatility.

This model is directly useful for developers building Aave-related infrastructure, especially in automation and risk monitoring. Liquidation bots can use the formula to prioritize positions based on short-term liquidation probability, reducing unnecessary checks and API calls. Off-chain dashboards and monitoring tools could show not only health factors but also estimated liquidation risks over time windows. This helps DeFi analysts and risk teams detect emerging threats across collateral pools without relying on simulation-heavy backends. Given its low computational cost, the formula fits well in lightweight scripts and browser-based trackers — from validator dashboards to liquidation queue managers.

\section{Conclusion}

In this paper, we derived and validated a closed-form analytical formula for computing liquidation probability in DeFi lending protocols. By modeling the evolution of cryptocurrency prices as a zero-drift geometric Brownian motion and applying the reflection principle and inverse Gaussian distribution from probability theory, we were able to express the likelihood of liquidation in a compact, non-simulation-based form. This method avoids the computational overhead of traditional Monte Carlo simulations while retaining strong accuracy across various time horizons. Through a thorough numerical comparison, we showed that the theoretical estimates match closely with empirical results from both Euler–Maruyama and exact GBM-based simulations, with error remaining stable and small.

Our results have direct practical implications. The analytical approach we propose enables real-time liquidation risk analysis, which can be particularly useful for protocol designers, liquidation bots, and institutional risk managers operating in volatile DeFi markets. It also provides a mathematically grounded foundation for designing low-latency monitoring systems that do not depend on high-throughput simulations. While the current model assumes zero drift, constant volatility and constant loan value, future extensions could incorporate stochastic volatility, user leverage, and dynamically adjusting liquidation thresholds. This work demonstrates how well-established mathematical principles can bring analytical clarity and computational efficiency to modern decentralized financial systems, contributing toward more robust and transparent risk assessment frameworks.

\section*{Author Contribution}

\noindent CRediT: Conceptualization: T.B., G.V.; Data curation: T.B., G.V.; Formal analysis: T.B.; Funding acquisition: No funding was received for this work; Investigation: T.B., G.V.; Methodology: T.B.; Project administration: T.B., G.V.; Resources: T.B., G.V.; Software: G.V.; Supervision: T.B., G.V.; Validation: T.B., G.V.; Visualization: T.B, G.V.; Writing - original draft: T.B.; Writing - review and editing: T.B., G.V.

\bibliography{sample}

\end{document}